\documentstyle[multicol,prl,aps,epsf]{revtex}

\begin{document}

\title{Dynamical Behaviour of Fine Granular Glass/Bronze Mixtures 
under Vertical Vibration}

\author{N. Burtally, P. J. King and Michael R. Swift}
\address{School of Physics and Astronomy, 
University of Nottingham, Nottingham NG7 2RD, 
England}

\maketitle

\begin{abstract}
We report the behaviour of mixtures of fine bronze and glass  
spheres under sinusoidal vertical vibration.  Depending upon the ratio of 
their diameters and the amplitude and frequency of the 
vibration, we observe the 
formation of sharp separation boundaries between glass-rich and bronze-rich 
phases, the absence of gross convection which would mix these phases, and 
a number of oscillatory and non-periodic behaviours.  These phenomena 
are related to the differential air damping of the glass and bronze grains,  
disappearing completely in the absence of air.


\end{abstract}

\begin{multicols}{2}

The dynamical behaviour of a shaken granular material is strongly influenced 
by the inelastic nature of collisions between the grains and with the container walls 
\cite{mac}.  Vertical sinusoidal vibration induces convection, surface waves, arching and 
pattern formation \cite{raj} depending principally upon the peak acceleration of the 
container, conveniently represented by $\Gamma = a\omega^2/g$.  
Here a is the peak amplitude of 
vibration, $\omega$ is the angular frequency and g is the acceleration due to gravity. 
There is also a substantial body of knowledge on the separation and segregation of 
granular mixtures under vertical vibration, but a clear understanding of many of 
the physical processes involved is still lacking \cite{ott}.  
Furthermore much of the effort 
has been concentrated on studying systems of large particulates for which air effects 
are unimportant. 

If a body of particulates containing a single larger and 
heavier grain is vibrated vertically, 
this grain tends to move to the upper surface, 
the ``Brazil nut effect''\cite{ros}; a larger lighter grain may under appropriate
circumstances move to the bottom, the ``reverse Brazil nut effect''\cite{shin}. 
Computer simulations, in which air 
effects are absent, also predict conditions for the ``Brazil nut'' to take up an
intermediate height\cite{shis}.
Others have simulated the behaviour 
of binary mixtures under vertical vibration and have offered predictions for separation 
based upon mass and size differences\cite{hong}.  None of these simulations include the 
interactions between the grains and the container walls that, in practice, induce 
convective stirring, and most of these predictions have yet to be tested experimentally. 

The important influence of air on fine grains was first 
recognised by Faraday\cite{far}.  Upon a vertically vibrated horizontal surface, the 
presence of air induces piling and modified convection in the granular bed.  Fine grains vibrated 
vertically within a container break symmetry by forming a tilted upper surface down 
which grains cascade\cite{far,eve}.   Faraday suggested that grains are 
sucked under the main body as the supporting platform accelerates downwards.  Later, 
the granular body falls upon these grains, causing continuously 
erupting piles\cite{far}.  However, while work at high $\Gamma$ supports this explanation\cite{tom},
other experiments show that 
it is not satisfactory for $\Gamma$ just
exceeding unity\cite{ind}, and a full explanation of the effects of the ambient fluid on the 
motion of fine grains is still the subject of debate\cite{tom,ind,dur}.

Here we report the influence of air on the behaviour of mixtures of fine bronze 
and soda-glass spheres under vertical sinusoidal vibration.  We observe the 
appearance of very sharp boundaries between glass-rich and bronze-rich phases and a 
range of periodic and non-periodic behaviours of the separation boundaries.  We also 
note the great disparity between the kinetic activity 
in the bronze-rich and in the glass-rich 
regions and the very high velocity shear which may exist across the separation boundaries.
	
Our experiments use bronze spheres of density $\rho_b = 8900 kg/m^3$ and soda-glass 
spheres of density $\rho_g = 2500 kg/m^3$, 
with four ratios of the mean diameters, $d_b/d_g$. The glass and 
bronze is sieved to produce a spread of sizes of typically $\approx \pm 10\%$ to avoid 
crystallisation.  The dynamic angles of repose lie within 
$23.4^o \pm 0.8^o$ and $23.9^o \pm 0.8^o$ 
for the glass and bronze spheres respectively and the coefficients of restitution, 
measured in vacuum, are both very close to unity.  A chosen mixture, of mean depth 
20mm, is contained within a rectangular soda-glass box 50mm high and of internal 
dimensions either 40mm x 10mm, 20mm x 10mm or 10mm x 10mm in the horizontal plane.  
The boxes are excited vertically in the frequency range $10Hz < f < 180Hz$ by an 
electromagnetic transducer, the axis of the transducer and the sides of the box being 
aligned to the vertical to within $1^o$.  The motion is monitored by cantilever 
capacitance accelerometers.  
The finest glass spheres, particularly those of diameters 
less than $100\mu m$, are influenced by static charge when shaken vigorously within a 
glass box.  The effects which we shall describe are still observed, 
but the glass spheres 
stick to the walls of the box, impairing photography.  The addition of minute 
quantities of an anti-static surfactant greatly reduces this effect.  
Initially we shall 
describe the behaviours of four mixtures contained in a 40mm x 10mm box.

\noindent
{\bf Mixture A}:  bronze $125-150\mu m$, glass $60-90\mu m$, 
$25\%:75\%$ by volume.   Figure 1(a) 
provides a simplified outline of the principal 
behaviours found at various values of f and $\Gamma$. 
At lower 
frequencies global convection and tilting of the upper surface are observed as the 
amplitude of vibration is increased above $\Gamma$ = 1.  
At the critical line shown in Fig. 1(a) as A, 
sharp separation boundaries quickly appear between glass-rich and bronze-rich 
regions.  The bronze-rich regions rapidly merge into a single {\em upper} bronze-rich 
phase which lies above a {\em lower} glass-rich phase.  
The boundary between bronze-rich and glass-rich phases is extremely well defined,
being only one grain-diameter wide.
The bronze-rich phase contains only a 
small proportion of glass $(5-20\%)$ 
while the lower region consists almost entirely of glass.  
Figure 2 shows this behaviour as a function of time for $\Gamma$ = 5.5 
and f=35Hz. Following the application of vibration, fine bronze-rich structures quickly appear.
Development towards a single upper bronze-rich region occurs by
coarsening, an effect observed in other granular systems\cite{mul}.
Once 
separation has happened convection currents occur within the individual bronze and 
glass-rich regions but do not act to cause mixing.   

\begin{figure}
\centerline{
\epsfysize=0.45\columnwidth{\epsfbox{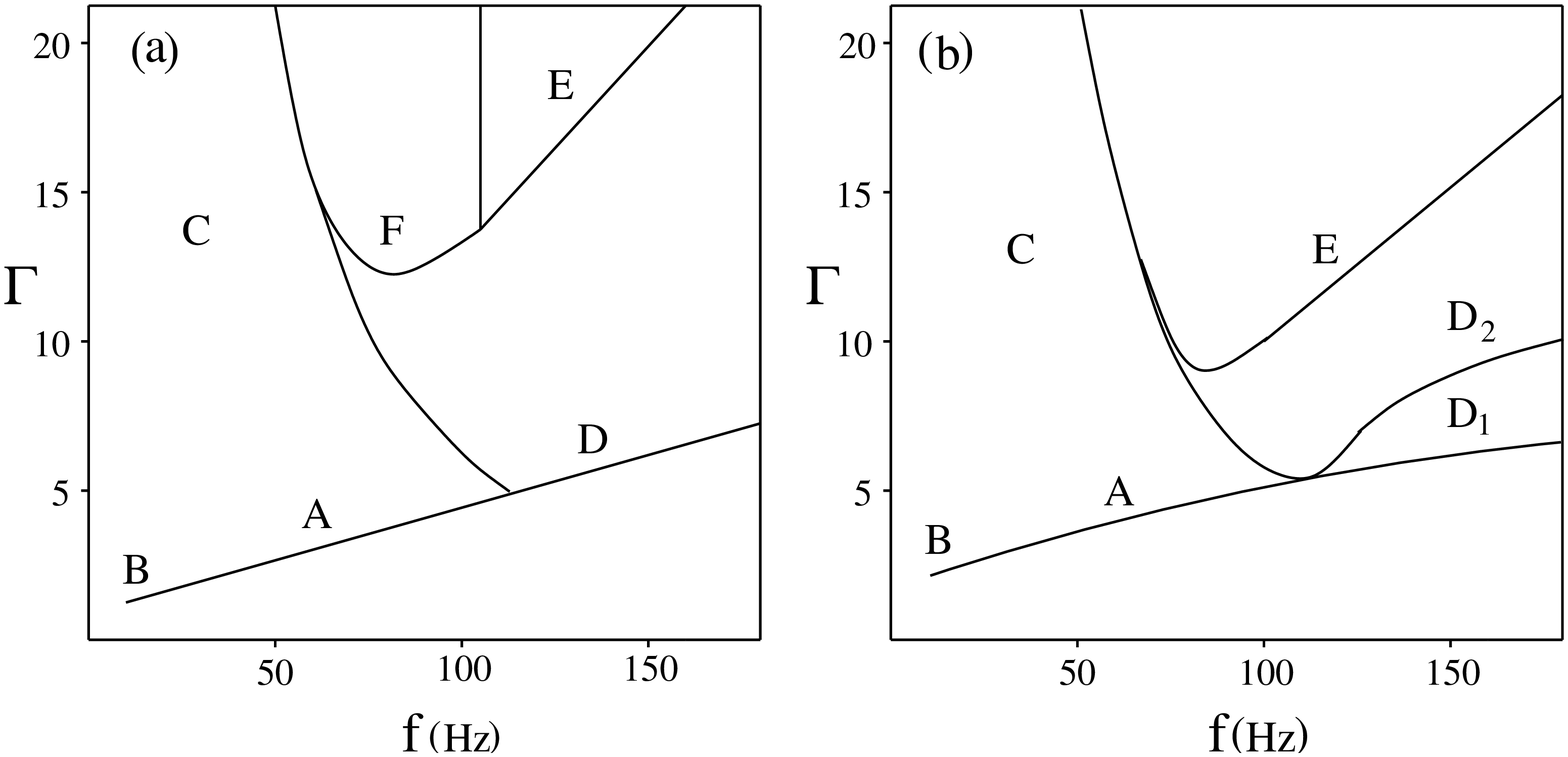}}}
\vskip 0.1 truein
\narrowtext{\caption{
The principal behaviours of mixture {\bf A} (a) and mixture {\bf B} (b)
as a function of f and $\Gamma$.} 
\label{Fig1}}
\end{figure}

Above A both 
the separation boundary and the upper 
surface exhibit fluctuations. At low values of $\Gamma$ and f,
region B in Fig. 1(a), simple oscillations back and forth between the two 
alternative tilts occur.  
At higher values of $\Gamma$ 
the fluctuations contain both periodic and 
non-periodic components.  By values of $\Gamma$ 
corresponding to region C, 
considerable throwing of the upper surface is observed and global convection currents 
act to mix bronze-rich and glass-rich regions; simple separation into two phases is no 
longer observed.

At higher frequencies there is also a critical line, D, at which sharp 
separation boundaries between a bronze-rich phase and a glass-rich phase first appear. 
Here, however, the bronze-rich regions rapidly merge to form a stable 
single layer at an intermediate height, sandwiched between upper and lower 
glass-rich regions.  The formation of such a sandwich is shown in Fig. 3 for 
$\Gamma$ = 6.8 and f =  
160Hz.  Again, the bronze-rich layer contains appreciable glass, but the glass-rich regions are 
almost completely free of bronze.  Convection may be observed within the 
individual glass and bronze-rich regions, with considerable velocity shear at the sharp 
separation boundaries, but no convection currents are present which would mix the 
two phases.  At a far higher critical line E, 
the system undergoes 
an inversion.  The bronze-rich layer rises slowly to the surface, while remaining  close to
horizontal.  Once there, the surface tilts.  The bronze then avalanches down the slope, 
and passes into the depths of the glass to form a stable low-lying horizontal layer. 
Over an intermediate range of frequencies (F in Fig. 1(a)) a similar process repeats 
continuously.  An intermediate bronze-rich layer rises to the surface, which tilts.  
The bronze-rich phase then passes into the depths to reform a horizontal layer which then 
begins to rise to the surface again. The sequence then repeats, oscillating between the two 
alternative tilts of the upper surface.  Such a sequence is shown in Fig. 4 for 
$\Gamma$ = 16.7
and f = 70Hz.

Both at low and at high frequencies, there is 
considerably more kinetic activity in the bronze-rich phase than in the 
glass-rich phase. ``Droplets'' of glass frequently condense within the bronze-rich 
phase, these droplets then moving to join nearby glass-rich regions.  In the high frequency 
regime we also observe small and violently vibrating elliptical excitations which 
traverse the upper horizontal surface of the bronze-rich layer when this layer is 
sandwiched between two glass-rich regions.  These disturbances appear to be regions of very 
low granular density.

\begin{figure}
\centerline{
\epsfysize=0.62\columnwidth{\epsfbox{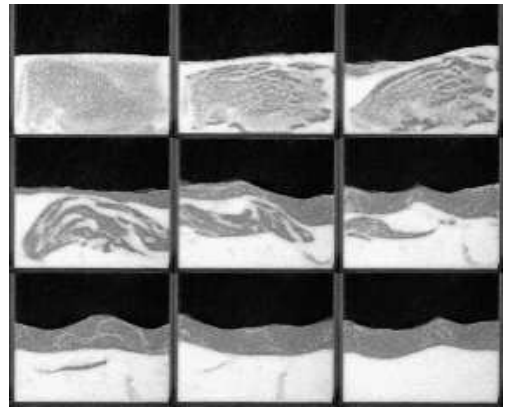}}}
\vskip 0.1 truein
\narrowtext{\caption{
Behaviour of mixture {\bf A} under $\Gamma= 5.5$ at 35 Hz showing the formation of
an upper bronze-rich region.  The pictures form a time sequence from upper
left to lower right. The fifth picture was taken after 16s, the seventh after
2mins and the ninth after 12mins.}
\label{Fig2}}
\end{figure}

\noindent
{\bf Mixture B}: bronze $90-120\mu m$, glass $90-120\mu m$, 
$25\%:75\%$ by volume. For this 
mixture we observe many similar phenomena to those described above\cite{bur}.  The 
behaviour again broadly falls into a low frequency regime and a high frequency 
regime (Fig. 1(b)).  In the low frequency regime we again observe the formation of a 
very sharp separation boundary, the bronze-rich region preferring to be above the 
glass-rich region (Fig. 1(b)-A).  We observe tilt oscillations, B, fluctuations of the 
separation boundary and upper surface, and at sufficiently high 
$\Gamma$, failure of 
separation by throwing and mixing convection, C.  In the higher frequency 
regime we observe the bronze-rich phase to settle at an intermediate height, with the 
sharp lower boundary, $D_1$, forming at somewhat lower values of $\Gamma$ than the upper 
boundary, $D_2$.  At a higher value of 
$\Gamma$, E, the bronze-rich phase moves to the upper 
surface which tilts; the bronze-rich phase then settling to a stable lower horizontal sandwich 
position. We observe upper glass-rich surfaces to be horizontal but for slight 
hexagonal patterning.  

\begin{figure}
\centerline{
\epsfysize=0.62\columnwidth{\epsfbox{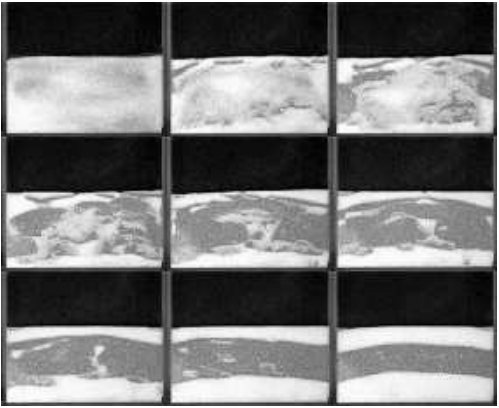}}}
\vskip 0.1 truein
\narrowtext{\caption{
Behaviour of mixture {\bf A} under $\Gamma= 6.8$ at 160 Hz showing the formation
of a glass/bronze/glass sandwich.  The pictures form a time sequence from upper left
to lower right. The fifth picture was taken after 50s and the ninth picture
after 7mins.}
\label{Fig3}}
\end{figure}

If the bronze-rich layer rises to the surface and passes again into the 
depths small amounts of bronze may be trapped upon the upper glass-rich surface where 
they collect together in a one or more swarms, vibrating vigorously in slight 
indentations in the surface formed by patterning and by their own weight.  These 
vibrating bronze puddles attract each other and may merge.  In doing so it often 
happens that the glass-rich surface can no longer sustain their weight.  Rather they form a 
droplet, which falls through the glass to join the lower bronze-rich layer.
For mixture {\bf B}, too, the kinetic activity within the glass-rich phase is far less than within 
the bronze-rich regions.   However, while the glass-rich phase contains a trace of  bronze, 
the bronze-rich 
phase contains considerably less glass than for mixture {\bf A}.  Correspondingly the 
condensation of glass ``droplets'' is far less obvious for mixture {\bf B} 
than for mixture {\bf A}.

\noindent
{\bf Mixture C}:  bronze $60-90\mu m$, glass spheres $125-150\mu m$, 
$25\%:75\%$ by volume. For this mixture we observe poor separation, with sharp boundaries only visible
over the limited frequency range of 25-90Hz. We observe the formation
of bronze-rich regions close to the upper surface at some lower frequencies and bronze-rich
regions at intermediate levels at some higher frequencies. We also observe
some oscillatory behaviours of the bronze-rich regions. However,
the separation is always very poor; the sharp separation boundary, where it does
exist, distinguishes regions containing a considerable proportion of the
other component. At all frequencies increasing $\Gamma$ readily induces global convection
currents which thwart any tendency to separate. 

\noindent
{\bf Mixture D}: bronze $38-63\mu m$, glass $125-150\mu m$, $25\% : 75\% $ by volume.  
For mixtures of 
these diameters we observe convective mixing for values of $\Gamma $ just greater than 1.  
At a value of 
$\Gamma$ in the range 1.8-2.6, weak separation is observed 
with glass predominating close 
to the upper surface and the lower regions remaining well mixed.   Diffuse 
rather than sharp separation boundaries are observed.  If 
$\Gamma$ is then slightly increased, 
the upper glass-rich region initially becomes purer.  Soon, however, the upper surface tilts and 
at the same time global convection currents appear.  These effectively mix the bronze 
and glass components and any separation disappears. 

\begin{figure}
\centerline{
\epsfysize=0.62\columnwidth{\epsfbox{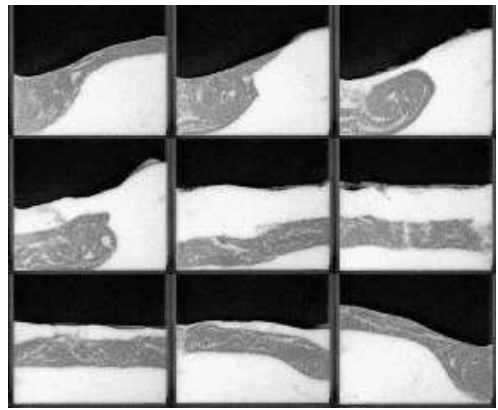}}}
\vskip 0.1 truein
\narrowtext{\caption{
Behaviour of mixture {\bf A} under $\Gamma= 16.7$ at 70 Hz showing one half period
of an inversion oscillation. The pictures form a time sequence from upper left
to lower right. The time period for a full oscillation is 30s.}
\label{Fig4}}
\end{figure}

In the experiments on mixtures {\bf A}-{\bf D} the behaviour is close to two dimensional; 
there is very little variation of any property across the smaller horizontal dimension.  
If the experiments are repeated in 20x10mm or 10x10mm boxes most of the same 
general separation phenomena are observed.  However, as the larger horizontal dimension 
is decreased, oscillations of the
separation boundaries are greatly reduced and the ability of 
bronze-rich layers to quickly pass to intermediate positions, by avalanching down an 
upper slope and passing en masse into the depths, is suppressed.  Instead, approach to 
equilibrium at an intermediate depth occurs by slow diffusion and droplet formation; 
it may become very slow indeed for the 10x10mm box geometry.

If, in any of the experiments, the box is evacuated separation effects cease 
once the pressure falls to below a few mbarr\cite{pak}.  Rather rapid global convection is 
observed which effectively maintains a near-homogeneous mixture.    

Our principal observation is the spontaneous formation of glass-rich and 
bronze-rich phases with very sharp separation boundaries.   
It is clear that air 
is responsible for this effect since it does not 
occur for large particles and it disappears 
if the box is evacuated to the point where the air viscosity 
is greatly reduced. A simple 
dynamical model which incorporates the effect of air viscosity through Stoke's law 
suggests that the relative acceleration of the glass and bronze spheres 
induced by air flow is given by 
$R=\rho_b d_b^2 /\rho_g d_g^2$.  R = 12, 3.6, 1 and 0.4 respectively for our mixtures {\bf A} to {\bf D}.  
R is large 
for the two mixtures which separate well with the bronze uppermost at low 
frequencies, and is somewhat less than unity for mixture {\bf D} which separates weakly
with the glass uppermost. 
Although Stoke's law is not expected to be accurately valid 
for our dense granular systems this simple argument
suggests that it is the differential air 
damping of the two species which is responsible for our observations. 

Our experiments suggest that if $R>>1$ there is a strong tendency for glass and bronze to
separate. In this limit, the bronze spheres will be only weakly affected
by the surrounding air, whereas the glass spheres will be strongly influenced by
the local air flow, which will tend to dampen the particles' motion. Thus, an
individual bronze sphere will be more mobile than a glass sphere and, consequently,
bronze-rich regions will be more dilated than glass-rich regions. 
Any initial inhomogeneity in the composition of the mixture will therefore set up number-density 
gradients under vibration. There will then be a tendency for particles to move
down these gradients, and, as the bronze spheres are more mobile, there
will be a net flux of {\em bronze} from the more compact to the less compact regions.
As the local concentration of bronze increases, this will be accompanied
by an increase in the dilation locally, thus enhancing the effect further. There
is therefore a dynamical instability that will induce separation of the two components.

Once well-defined bronze-rich and glass-rich regions have formed,
it will be extremely difficult for a bronze sphere to re-enter the dense glass-
rich region, rather it will ``bounce off''.  
Thus, the glass-rich regions will naturally tend to lose any bronze component, and eventually
consist almost entirely of glass, as we observe.   On the other hand, it is easier for a glass 
sphere to enter a dilated bronze-rich region.  For example a high-energy bronze 
sphere, bouncing off the interface with the glass-rich region may, in the process, 
knock a glass sphere into the bronze-rich region.  The bronze-rich regions will 
therefore contain a proportion of glass spheres as is observed. As R is increased, the spaces between 
the bronze spheres available to glass spheres will increase and, 
one would expect an increased 
proportion of glass within the bronze-rich regions for increased R, as we also observe.    

However, the above discussion only provides a qualitative picture of the separation
phenomena and a satisfactory quantitative model is still lacking.
Such a model will also have to explain the preferred position of the 
bronze-rich region.
Two effects may contribute. The 
less highly damped component will tend to be thrown higher, leading to bronze being 
uppermost for $R >>1$ and glass being uppermost for $R << 1$.  However, once 
separation has occurred, the dilation, and thus 
the mean density, of the two phases will differ.  The equilibrium position of the 
phases will then be influenced by their relative buoyancy, which in turn is determined 
by the densities of the phases and by their dilation.  
The dilation is itself a function 
of f, $\Gamma$ and the position within the granular body.  We suggest that at higher 
frequencies the mean buoyancy has a major influence upon the relative positions of 
the two phases.

We are grateful to T. Davies, N. Miles and S. Kingman for their assistance, and
to Potters Ballotine Ltd for advice and samples.

\end{multicols}
\end{document}